\documentstyle[12pt]{article}

\def\non{\nonumber}
\def\lab{\label}

\def\be{\begin{equation}}
\def\ee{\end{equation}}
\def\bea{\begin{eqnarray}}
\def\eea{\end{eqnarray}}

\def\eg{{\it e.g.\ }}

\def\dag{\dagger}
\def\ra{\rightarrow}
\def\del{\partial}
\def\lsim{\mbox{{\scriptsize \raisebox{-.9ex}
      {$\;\stackrel{{\textstyle <}}{\sim}\,$} }} }

\def\calL{{\cal L}}\def\calO{{\cal O}}
\def\bfq{{\bf q}}
\newcommand{\e}{{\mbox{e}}}
\def\CPT{{\small $\chi$PT}}
\def\cLHB{\calL^{\mbox{\scriptsize HB}}_{\rm ch}}

\def\mN{m_{\mbox{\tiny N}}}

\begin{document}

\hspace*{11cm}USC(NT)-Report-01-5

\vspace{4mm}
\noindent
{\bf ELECTROWEAK PROBES AND NUCLEAR EFFECTIVE 
FIELD THEORY}\footnote{Invited talk at International
Workshop on Physics with GeV Electrons and Gamma-Rays,
Sendai, Japan, February 2001}

\vspace{1cm}
\noindent
{\bf K. Kubodera}\footnote{Supported in part by 
the U.S. National Science Foundation, 
Grant No. PHY-9900756}\\
Department of Physics and Astronomy, 
University of South Carolina, \\ 
Columbia, South Carolina 29208, USA\\
 
\vspace{0.5cm}       
\noindent
{\bf Abstract:}
There is growing interest in nuclear physics 
applications of effective field theory. 
I give a brief account of some of the latest developments
in this area.  I also describe interplay between
this new approach and the traditional nuclear physics approach.

\vspace{4mm}

As is well known, the description of nuclear structure
based on the phenomenological potential picture
has been extremely successful \cite{cr}.
In this picture, nuclear responses 
to external probes are given by one-body
impulse approximation terms and
exchange-current terms (usually two-body terms)
acting on nuclear wave functions,
with the exchange currents derived
from a one-boson exchange model.
In a modern realization of this approach \cite{crit},
the vertices characterizing relevant Feynman diagrams
are determined with the use of the low-energy theorems
and current algebra.
We may refer to this type of formalism as SNPA, 
the {\it standard nuclear physics approach}.
SNPA has been used extensively 
for nuclear electroweak processes, and
the reported good agreement between theory and experiment
suggests the basic soundness of SNPA \cite{cr}.
Meanwhile, one of the major challenges 
in nuclear physics today is to establish a connection 
between nuclear dynamics and the fundamental QCD.
Effective field theory (EFT) \cite{wei79,gl84}
is considered to offer 
a natural and useful framework for this purpose.
In this talk I first give a highly abridged account 
of EFT as applied to nuclear physics.
I then discuss what may be called 
a ``symbiotic" relation
between EFT and SNPA.

The basic idea underlying EFT is quite simple.
In describing phenomena
characterized by a typical energy-momentum scale $Q$,
we need not explicitly include
in our Lagrangian those degrees of freedom 
which are characterized by energy-momentum
scales much larger than $Q$.
Then, introducing a cut-off scale $\Lambda$
that is substantially larger than $Q$,
we may classify our fields (generically denoted by $\Phi$)
into two parts: a high-energy part $\Phi_{\mbox{\tiny H}}$
and a low-energy part $\Phi_{\mbox{\tiny L}}$.
{\it Integrating out} $\Phi_{\mbox{\tiny H}}$
leads to an {\it effective} Lagrangian 
that only involves $\Phi_{\mbox{\tiny L}}$;  
thus the original Lagrangian 
${\cal L}$ and the effective Lagrangian 
${\cal L}_{{\rm eff}}$ are related to each other as
\be
\int\![d\Phi]\e^{{\rm i}\int d^4x{\cal L}(\Phi)}
= \int\![d\Phi_{\mbox{\tiny L}}]
\e^{{\rm i}\int d^4x{\cal L}_{\rm eff}
(\Phi_{\mbox{\tiny L}})}
\ee
${\cal L}_{{\rm eff}}$ inherits all the symmetries 
(and the patterns of symmetry breaking, if any)
of the original ${\cal L}$,
and therefore ${\cal L}_{{\rm eff}}$
should consist of all possible monomials of 
$\Phi_{\mbox{\tiny L}}$ and their derivatives
that are consistent with the symmetry requirements.
Since a term involving $n$ derivatives
is characterized by $(Q/\Lambda)^n$,
we have perturbative expansion
in terms of a small parameter $Q/\Lambda$.
The coefficients of the monomials appearing in 
this expansion are called
the low-energy constants (LEC).
Provided all the LEC's up to a specified order are known,
we have a complete (and hence model-independent) Lagrangian.
In a nuclear-physics application of EFT,
the original Lagrangian ${\cal L}$ is the QCD Lagrangian
while, in the energy-momentum regime
$Q\ll \Lambda_{\mbox{\tiny QCD}}\sim 1$ GeV,
${\cal L}_{{\rm eff}}$ is expected to
involve hadrons rather than quarks and gluons.
Furthermore, for $Q\lsim m_\pi$,
it is expected that we only need the pions and nucleons
as explicit degrees of freedom.
However, chiral symmetry of QCD (and its small violation
due to the quark mass terms)
must be inherited by our ${\cal L}_{{\rm eff}}$.
An EFT constructed this way is called 
chiral perturbation theory ($\chi$PT) \cite{gl84,wei90}.
A problem one encounters in including the nucleon in $\chi$PT
is that the nucleon mass $\mN$ is comparable to 
$\Lambda_{\mbox{\tiny QCD}}$, but
the heavy-baryon chiral perturbation theory
(HB$\chi$PT) allows us to avoid this difficulty.
In HB$\chi$PT, effective degrees of freedom
are the pions and the large components of nucleons (denoted by $B$),
and $\cLHB$ has as its expansion parameters
$Q/\Lambda_{\mbox{\tiny QCD}}$, 
$m_\pi/\Lambda_{\mbox{\tiny QCD}}$
and $Q/\mN$.
Since $\mN\approx \Lambda_{\mbox{\tiny QCD}}$,
we may characterize chiral and heavy-baryon expansions
jointly by introducing the index ${\bar \nu}$ 
defined by $\bar{\nu}=d+(n/2)-2$,
where $n$ is the number of 
fermion lines belonging to a vertex,
and $d$ is the number of derivatives
(with $m_\pi$ counted as one derivative). 
For instance, the leading-order ($\nu=0$) term
is given by 
\bea
{\cal L}^{(0)}&= &
 \frac{f^2_\pi}{4} \mbox{Tr} 
[ \partial_\mu U^\dagger \partial^\mu U 
 + m_\pi^2 (U^\dagger +  U - 2) ] \non
\\ &&
 + \bar{B} ( i v \cdot D + g_A^{} S \cdot u ) B 
 - \frac12 \displaystyle \sum_A 
      C_A (\bar{B} \Gamma_A B)^2 
\lab{eq17a}
\eea
Here $U(x)$ is an SU(2) matrix field
related non-linearly to the pion field,
$\xi\equiv\sqrt{U}$,
$u_\mu \equiv i (\xi^\dag \del_\mu \xi
                - \xi \del_\mu \xi^\dag)$, 
$S_\mu=i\gamma_5\sigma_{\mu\nu}v^\nu/2$, and 
$D_\mu$ is the covariant derivative acting on the nucleon.
According to Weinberg's counting rule \cite{wei90},
a Feynman diagram consisting of $N_A$ nucleons,
$N_E$ external fields,
$L$ loops and $N_C$ disjoint parts
is of $\calO(Q^\nu)$ with
$\nu = 2 L + 2 (N_C-1) + 2 - (N_A+N_E) + \sum_i \bar \nu_i$.
This counting scheme works well
for cases with zero or one nucleon.
For nuclear systems ($A\ge 2$), however, 
its straightforward application fails 
because our systems involve very low-lying excited states,
which spoil the ordinary counting rule \cite{wei90}.
To avoid this difficulty,
it is useful to classify Feynman diagrams into two categories.
Diagrams in which every intermediate state
contains at least one meson in flight
are classified as irreducible diagrams,
and all others as reducible diagrams.
Following Weinberg, we apply the chiral counting
only to irreducible diagrams.
The sum of all the irreducible diagrams 
(up to a specified chiral order)
is then used as an effective operator
residing in the nuclear Hilbert space.
Summing up infinite series of irreducible diagrams
(solving either the Schr\"odinger equation 
or the Lippman-Schwinger equation)
incorporates the contributions 
of reducible diagrams \cite{wei90}.
We refer to this two-step procedure as
{\it nuclear} \CPT\ .

An alternative counting scheme,
called the polynomial divergence subtraction (PDS),
was proposed by Kaplan, Savage and Wise \cite{ksw},
and it has been used extensively to calculate 
many observables in the two-nucleon systems.
PDS has the advantage
that it formally preserves chiral invariance,
whereas the Weinberg scheme loses
manifest chiral invariance.
This problem, however, does not occur
up to the chiral order of our concern,
i.e., the chiral order up to which our irreducible diagrams
are calculated.\footnote{
For a detailed discussion of the relation between
PDS and the Weinberg scheme, see a review by
Brown and Rho \cite{kol}.}

The past several years have witnessed 
an explosion of nuclear \CPT\  work in 
both the Weinberg scheme and PDS;
for a recent review, see \eg \cite{kol}.
Here I talk mostly of the Weinberg approach,
with of course no intention to minimize 
the achievements made in PDS
(see, Beane et al. \cite{kol}).

In applying nuclear \CPT\ to cases that involve 
external probes,
we equate a nuclear transition operator ${\cal T}$ 
to the set of all the irreducible diagrams
(up to a given order $\nu$) 
with the external current inserted.
To maintain formal consistency, 
${\cal T}$ should be inserted 
between the initial and final 
nuclear states that are governed
by the nuclear interactions that subsume
all irreducible A-nucleon diagrams 
of up to $\nu$-th order. 
If this program is fully carried out,
we would have an {\it ab initio} calculation.
In practice, however, 
we often use initial and final nuclear wave functions 
obtained with the use of  
phenomenological nucleon-nucleon interactions.
This eclectic method may be called 
a {\it hybrid approach} to nuclear \CPT.
The hybrid approach has proved highly successful
in many cases.  Just to mention one example,
a hybrid HB$\chi$PT calculation
for the isovector M1 transition amplitude
in the $n({\rm thermal})+p\ra d+\gamma$ reaction
\cite{pmr95} was carried out by Park et al. \cite{pmr95},
and the results exhibited perfect agreement
with the data.

Although the hybrid approach works very well,
it is desirable to formally justify this approximation.
For the two-body systems,
Park, Min, Rho and myself (PKMR) \cite{pkmr} 
have carried out an {\it ab initio} calculation 
up to next-to-leading-order (NLO).
Because there are no loop contributions up to NLO,
we can simply work with a generic potential of the form
\bea
{\cal V}(\bfq) &=&
- \tau_1\cdot\tau_2 \,\frac{g_A^2}{4 f_\pi^2}\,
\frac{ \sigma_1\cdot \bfq\,\sigma_2\cdot \bfq}{\bfq^2+m_\pi^2}
+ \frac{4\pi}{m_{\mbox{\scriptsize N}}}
 \left[C_0 + (C_2 \delta ^{ij} + D_2 \sigma^{ij})
 q^i q^j \right] \label{Vq}
\eea
where
$\sigma^{ij} = 3/\sqrt{8}[
(\sigma_1^i \sigma_2^j + \sigma_1^j \sigma_2^i)/2
- (\delta^{ij}/3) \sigma_1 \cdot \sigma_2 ],$
with $\bfq$ the momentum transfer.
The first term coming from 
Goldstone-boson (pion) exchange 
is uniquely determined by \CPT.
The (local) terms in the square brackets
encode the effects of those degrees of freedom 
that have been integrated out.
Inserting the above potential 
into the Lippman-Schwinger equation
gives rise to an infinite series of divergences,
which may be regularized 
by a momentum-cutoff parameter, $\Lambda$.
For a given value of $\Lambda$
the parameters $C$'s and $D$'s
are determined by relating them (after renormalization)
to the low-energy two-nucleon observables;
the scattering length and the effective range 
for the scattering channels
or to a selected set of the deuteron observables.
Then predictions can be made for 
the N-N scattering phase shifts 
and the low-energy properties of the deuteron
(other than those used as input).
We can also make the {\it parameter-free} estimation
of electroweak observables.
According to PKMR's {\it ab initio} calculation, 
(1) all the calculated quantities 
are in good agreement with the empirical information;
(2) the results are stable against 
the variation of the cut-off parameter $\Lambda$,
so long as $\Lambda$ lies within a reasonable range;
this {\it reasonable} range is found to be
$\Lambda=$100\,-\,300 MeV 
($\Lambda=$200\,-\,500 MeV)
in the absence (presence) of the pion in the theory,
and this behavior is consistent 
with the general tenet of EFT.
(3) Good agreement between the {\it ab initio} calculation 
and the hybrid approach indicates that the latter 
is justifiable.
Now, the first question you could ask about
the relation between EFT and SNPA is:
Is there any real difference between EFT (as used here)
and the familiar effective-range formula ?
As a matter of fact, when our EFT is devoid of pions,
the two LEC's ($C_0$ and $C_2$) in the above expression
play essentially the same role as 
the effective-range expansion parameters,
$a$ and $r_e$.
But, with introduction of the pion,
there is no longer such trivial correspondence,
and \CPT\ expansion contains more physics in it.
In addition, \CPT\ offers
a unified expansion scheme 
for both N-N scattering
and nuclear transition processes.

The above example is concerned with observables
that are governed by the lowest order contributions,
but these are not the only cases for which
\CPT\ is useful.
To illustrate this point,
let me discuss the spin observables in 
$\stackrel{\ra}{n}+\stackrel{\ra}{p}\ra d+\gamma$
at thermal energy.
The photon circular polarization,
$P_\gamma$, and the photon anisotropy
are found to be highly sensitive to the small
isoscalar M1 and E2 transition matrix elements,
and that these matrix elements are controlled 
by terms of high chiral indices.
Recent two \CPT\ calculations
for these spin observables \cite{crs99}
agree with the existing experimental value of $P_\gamma$.
The important point is that, 
although higher-order calculations can in general lead to 
a plethora of unknown LEC's,
there are cases in which the physical amplitudes
depend only on limited combinations of LEC's
so that \CPT\ retains its predictive power.

Next I would like to touch upon the interface 
(as of today) between EFT and SNPA,
using as an example the weak-interaction processes 
in the two-body systems.
Calculations based on SNPA have been done 
for $\mu$-d capture
\cite{TKK}, $\nu$-d reactions \cite{TKK,NSGK} 
and pp-fusion \cite{crsw,pkmr}.
For the first two processes, 
for which experimental data are available,
the SNPA calculations reproduce the data well 
within the somewhat large
($\approx$10 \%) experimental errors.
How does EFT fare in this regard ?
Can we see any interplay between SNPA and EFT ?
Let me discuss this point using the $\nu-d$ reactions,
which are of great current importance
because of highly consequential experiments 
in progress at the Sudbury Neutrino Observatory 
(SNO) \cite{SNO}.
A primary goal of SNO is to study 
the solar neutrinos by monitoring
reactions occurring in heavy water:
$\nu_e+d  \rightarrow e^- + p + p$,
$\nu_x + d \rightarrow \nu_x + p + n \;\;
  (x=e,\,\mu,\,{\rm or}\,\,\tau)$,
$\bar{\nu}_e + d \rightarrow e^+ + n + n$,
and
$\bar{\nu}_x + d \rightarrow \bar{\nu}_x + p + n \;\;
  (x=e,\,\mu,\,{\rm or}\,\,\tau)$.
SNO, which can detect the charged-current (CC)
and neutral-current (NC) reactions 
separately but simultaneously,
provides valuable information
about neutrino oscillations.
Obviously, the accurate knowledge of the $\nu$-$d$ reaction 
cross sections is crucially important 
in interpreting the SNO experiments.
The most elaborate SNPA calculations
have been carried out very recently by Nakamura, Sato,
Gudkov and myself (NSGK) \cite{NSGK}.
The exchange currents in SNPA turn out to be
controlled essentially by the coupling constant
$g_{\pi N\Delta}$.
NSGK used for $g_{\pi N\Delta}$
the value given by the existing model
and the value adjusted to reproduce 
the tritium $\beta$-decay rate.
The difference between these two cases was adopted 
as a measure of theoretical uncertainty.
Meanwhile, an EFT calculation of the $\nu d$ cross sections 
has recently been reported by 
Butler, Cheng and Kong \cite{EFT}.
Their results agree with those of SNPA in the following sense.
As mentioned, some LEC's in EFT expansion often cannot be fixed 
by the symmetry requirements alone and hence 
need to be determined empirically. 
In \cite{EFT}, the coefficient $L_{\rm 1A}$ of
a four-nucleon axial-current counter term
enters as an unknown parameter,
although {\it naturalness} arguments suggest
$|L_{\rm 1A}|\le 6\,{\rm fm}^3$.
Butler et al. determined $L_{\rm 1A}$
by requiring that their EFT results reproduce
the cross sections of NSGK.
Remarkably, the adjustment of one parameter, $L_{\rm 1A}$,
leads to perfect agreement between
the cross sections obtained in \cite{EFT}
and those of NSGK
for all the four reactions under consideration.
The best-fit value of $L_{\rm 1A}$ 
is $L_{\rm 1A}=5.6\,{\rm fm}^3$ \cite{EFT},
a value quite reasonable compared 
with the dimensionally expected value. 
The fact that an {\it ab initio} 
calculation (modulo one free parameter)
based on EFT is consistent with the results of SNPA
renders strong support for the basic reliability of SNPA.
Thus EFT and SNPA are playing complementary roles here.
EFT, being a general framework, is capable of giving 
model-independent results, \underbar{\it provided}
all the LEC's in ${\cal{L}}_{eff}$ are predetermined. 
At present, however, ${\cal{L}}_{eff}$ does 
contain an unknown LEC, $L_{\rm 1A}$.
Meanwhile, although SNPA involves some elements
of models,
its basic idea and the parameters contained in it
have been tested with many observables.
Thus, insofar as the validity of these tests
is accepted, SNPA has predictive power.
We should also mention that,
beyond the solar neutrino energy regime,
SNPA is at present the only available formalism
for calculating the $\nu$-$d$ cross sections.
The EFT calculation in \cite{EFT},
by design, ``integrates out" all the degrees of freedom
but that of the heavy baryon.
The nature of this so-called ``nucleon-only" EFT 
limits its applicability to very low incident neutrino energies 
(typically the solar neutrino energies).
On the other hand, there is no obvious conceptual obstacle
in applying SNPA to an energy regime significantly higher
than that of the solar neutrinos.

Going back to EFT itself,
I must emphasize the importance
of the direct determination 
of the crucial LEC, $L_{1A}$ in the notation of
\cite{EFT} and $\hat{d}_R$ in the language of
\cite{pkmr}.
One possibility is to use 
$\mu^-+d\rightarrow \nu_\mu+n+n$.
The rather large energy transfer
due to the disappearance of $\mu^-$
may look worrisome but, since $\nu_\mu$
carries away most of the energy,
the process is in fact  a rather {\it gentle} one
as far as the hadron sector is concerned.
So, provided the quality of experimental data 
improves sufficiently,
$\mu$\,-\,$d$ capture can be useful to determined the LEC.
Another possibility is to use the tritium $\beta$-decay rate. 
This input was so far used in the context of SNPA
\cite{crsw,NSGK} 
but it can also be used to determine our LEC.

\end{document}